\documentclass[twocolumn]{aastex63}

\usepackage{amssymb}
\usepackage{amsmath}
\usepackage{graphicx}
\usepackage{dcolumn}
\usepackage{hyperref}
\usepackage{color,units}
\usepackage{aas_macros}
\usepackage{lineno}
\usepackage{xspace}
\usepackage{subfigure}
\usepackage{comment}
\usepackage{newtxtext,newtxmath}
\usepackage{natbib}

\usepackage[normalem]{ulem} 

\newcommand{\Msun}{\,{\rm M_\odot}}
\newcommand{\Mblack}{M_\bullet}
\newcommand{\vecb}[1]{\boldsymbol{#1}}

\usepackage{etoolbox}

\makeatletter

\patchcmd{\NAT@test}{\else \NAT@nm}{\else \NAT@nmfmt{\NAT@nm}}{}{}

\DeclareRobustCommand\citepos
  {\begingroup
   \let\NAT@nmfmt\NAT@posfmt
   \NAT@swafalse\let\NAT@ctype\z@\NAT@partrue
   \@ifstar{\NAT@fulltrue\NAT@citetp}{\NAT@fullfalse\NAT@citetp}}

\let\NAT@orig@nmfmt\NAT@nmfmt
\def\NAT@posfmt#1{\NAT@orig@nmfmt{#1's}}

\newcommand{\BHI}{Black Hole Initiative, Harvard University, Cambridge, MA 02138, USA}
\newcommand{\CFA}{Center for Astrophysics $\vert$ Harvard \& Smithsonian, Cambridge, MA 02138, USA}

\begin{document}

\title
{
    Detection Prospects of Local Super-Massive Black Holes Based on the Sloan-Digital Sky Survey
}

\author[0000-0002-4831-2105]{Nadav Joseph Outmezguine}
    \email{Nadav.Out@gmail.com}
\affiliation{School of Physics and Astronomy, Tel-Aviv University, Tel-Aviv 69978, Israel}
\affiliation{Berkeley Center for Theoretical Physics, Department of Physics, University of California, Berkeley, CA 94720, U.S.A.}
\affiliation{Theoretical Physics Group, Lawrence Berkeley National Laboratory, Berkeley, CA 94720, U.S.A}

\author[0000-0001-9879-7780]{Fabio Pacucci}
\affiliation{\CFA}
\affiliation{\BHI}

\author[0000-0003-4330-287X]{Abraham Loeb}
\affiliation{\CFA}
\affiliation{\BHI}

\begin{abstract}
\noindent We use the Sloan-Digital Sky Survey quasar catalog to statistically infer the local abundance of black holes heavier than $10^8\Msun$,  which allows us to estimate the detection prospect of super-massive black holes by future observational campaigns.  We find that the upcoming James Webb Space Telescope (JWST) and the Extremely Large Telescope (ELT) should be able to resolve, with integral field spectroscopy techniques, the gravitational influence of $\sim10^3$ black holes within a sphere of $\sim50\;\rm Mpc$. A Very-Long Baseline (VLB) observatory with one receiver placed in a geostationary orbit, is predicted to capture $\sim10$ images of the silhouette of a black hole, similar to the image of $\rm M87^*$ recently performed by the Event Horizon Telescope.
\end{abstract}
\keywords{Astrophysical black holes (98), Supermassive black holes (1663), Quasars (1319), Very long baseline interferometers (1768), Space telescopes (1547)}

\section{Introduction}
Super-massive black holes (BHs) are at the top of a broad BH mass distribution, with BHs mass ranging across roughly ten orders of magnitude.
In order to reach such an extreme variety in mass, BHs need to grow from their original seeds, formed at $z\sim 20-30$ \citep{2013fgu..book.....L}, via gas accretion and mergers \citep{Pacucci_2020}. 

Super-massive BHs are very rare, 
those with mass $\sim 10^{10} \Msun$ are expected to be $\sim 1000$ times more scarce than those with mass $\sim 10^6\Msun$ \citep{Habouzit_2020}. Their abundance also decreases significantly with redshift and super-massive BHs at $z\sim 0$ are roughly a $100$ times more prevalent than at $z\sim 4$ \citep{Cao_2010, Ueda_2014, Habouzit_2020}.

Currently the most massive BH observed, labeled TON 618, has a mass $\approx 6.6 \times 10^{10} \Msun$ estimated from the H$\beta$ emission line \citep{Shemmer_2004}. So far, only about $30$ BH at the $\sim 10^{10} \Msun$ scale were observed. The mass of those BHs is surprisingly close to the theoretical maximal mass predicted for a BH growing by luminous accretion of matter \citep{King_2016, Inayoshi_2016}. It is important to note that this mass limit, around $\sim 5\times 10^{10} \Msun$ with some variations depending on the accretion physics, is valid only for BHs that grow predominantly by active accretion of gas from an accretion disk. There is no theoretical upper bound on the mass of a BH that grows by mergers (e.g., \citealt{King_2016}, see also \citealt{Pacucci_2017, Woods_2019} for maximum-mass considerations in the high-$z$ Universe).

The presence of a BH in a given volume of space can be inferred using several methods. Gas accretion from a stable accretion disk (and corona) emits copious amounts of radiation, which can then be observed in a very broad range of frequencies in the electromagnetic spectrum, typically from radio frequencies all the way to hard X-rays. Local super-massive BHs are, however, challenging to detect via their emitted radiation, as most of them are no longer in an active accretion phase, due to a significant decrease in available cold gas at low redshift \citep{Power_2010}. In fact, the X-ray luminosity emitted by ultra-massive BHs was shown to decrease significantly in X-ray surveys at $z < 1$ compared to those at higher redshift \citep{Ueda_2014}.

Gravitational waves, with the Laser Interferometer Space Antenna (LISA) to play a fundamental role in the next decade, can also shed light on the presence of BHs. The mergers of BHs can emit a large amount of energy in the form of gravitational waves. Remarkably,  heavy super-massive BHs mergers will be challenging to detect even with LISA, due to a decreased sensitivity at masses larger than $\sim 10^9 \Msun$ \citep{LISA_2017}.

A BH can also have significant gravitational influence on nearby objects. In fact, the effect of the gravitational field of a BH on nearby stars has been spectacularly used to detect and accurately determine the mass of a super-massive BH at the center of our own galaxy \citep{Genzel_2003, Ghez_2008}. Furthermore, in a pioneering work, \cite{Kormendy_1995} applied for the first time dynamical search techniques based on the effects of BH on nearby gas and stars, to central BHs in 7 nearby galaxies, including M31, M32 and M87. Since then, the mass of many super-massive BHs have been measured dynamically using integral-field spectroscopy (IFS) techniques (see, e.g., \citealt{McConnell_2012} and the more recent \citealt{Liepold_2020}). Additionally, the gravitational field of a BH can produce gravitational lensing on the light of background sources \citep{Paczynski_1996}. 

Very recently, the observational techniques at our disposal to study BHs have been extended by the Event Horizon Telescope (EHT) collaboration, which used Very-Long Baseline (VLB) interferometry to study the shadow of a BH on scales comparable to its Schwarzschild radius \citep[][hereafter EHT19]{EHT}.
\defcitealias{EHT}{EHT19}

In this study we focus on super-massive BHs heavier than $10^8 \Msun$. We infer their local abundance from the statistics of quasars derived from the SDSS DR7~\citep{Schneider_2010, Shen_2011}, togther with the work ~\citet{2010MNRAS.406.1959S}, which calculated the fraction of black holes that shine as optically luminous quasars at a given time, denoted by $f_{\rm opt}$. This allows us to investigate the local abundance  of super-massive BHs as a function of the angular size of their shadows~\citep{1966MNRAS.131..463S,1979A&A....75..228L}, or of their gravitational influence radii \citep{Peebles_1972}.
This information enables us to predict how likely will it be for current and future VLB observatories to resolve a BH shadow image similar to that of $\rm M87^*$ \citepalias{EHT}. It also allows us to predict the number of BHs that could indirectly be observed via their gravitational effect on nearby stars and gas, via IFS with the James Webb Space Telescope (JWST) and, in the more distant future, with the Extremely Large Telescope (ELT).

\section{Data} 
\label{sec:data}
We restrict our attention to SDSS quasars with virial mass $\Mblack$ estimated as $\Mblack > 10^8 \Msun$, following \citet{Shen_2011}. {Note that virial masses in this database  may be over-estimated by a factor $\sim 3$ towards the high-end of the mass distribution (\citealt{Shen_2008}, see also Appendix~\ref{sec:effect_of_sdss_bias} for the effect on our results)}.  We will make use of \citepos{2010MNRAS.406.1959S} measurement of $f_{\rm opt}$ which is a redshift dependent quantity, for consistancy with their analysis we restrict our analysis to the same redshift interval as they did, $0.4\leq z\leq2.5$.  To simplify our analysis (see Sec. \ref{sec:methods} for further details) we seek the largest (co-moving) sphere that can reside entirely inside this redshift interval, within which the number density SDSS quasars is distributed uniformly. We find this sphere to have a co-moving radius of  $2300\,\rm{Mpc}$ and to be centred at $z=1.3$, $\mathrm{RA}=180$ degrees and $\mathrm{Dec}=32.4$ degrees. These cuts on the data leaves us with a total of $50478$ quasars and an observed co-moving volume of $\sim5.1\times10^{10}\;\rm Mpc^3$, resulting in a quasar-quasar co-moving mean separation of $\sim 62\,\rm Mpc$. This mean co-moving distance between quasars is consistent with results from previous high-$z$ quasar surveys (see, e.g., \citealt{Fan_2001} and \citealt{Pacucci_2019} about how this distance could be over-estimated).

\section{Method}
\label{sec:methods}
We assume the location of Earth in the universe is arbitrary. As such, the distribution of BHs seen by an observer located on Earth should be equivalent to that seen from any random point within the SDSS observed volume. Our methodology therefore proceeds as follows; We calculate the quasar mass function (QMF) seen from a random point in the SDSS observed volume and interpret it as an estimate of the QMF at $z\sim1.4$. Then, correcting for the fraction of halos that host luminous quasars, $f_{\rm opt}$, we obtain an estimate for the BH mass function (BHMF) at the same redshift. Bearing in mind that the activity of AGNs drops rapidly after $z\lesssim1$~\citep{Ueda_2014}, we treat the this BHMF a faithful estimate of the local ($z\sim0$) BHMF. Integrating the resulting BHMF over a certain range of BH mass and/or distance, we then predict the total number of  super-massive BHs with an angular size larger then a certain value. Using those results we predict the number of BHs a current or future observational campaign may find, per mass interval, distance interval and as a function of resolution. We note here that the contribution of BHs with a mass lower than roughly $10^9 \Msun$ to our analysis is under-estimated. This is the due to an incomplete selection of BHs resulting from  flux limitations (see.~\citealt{2006AJ....131.2766R,2009ApJ...699..800V,2010ApJ...719.1315K}.) Those BHS however do not drastically effect our analysis since their angular size is typically too small to be observed in the techniques we discuss.

\subsection{Estimating the Local Black Hole Mass Function}

\begin{figure}
\begin{center}
    \includegraphics[width=1\linewidth]{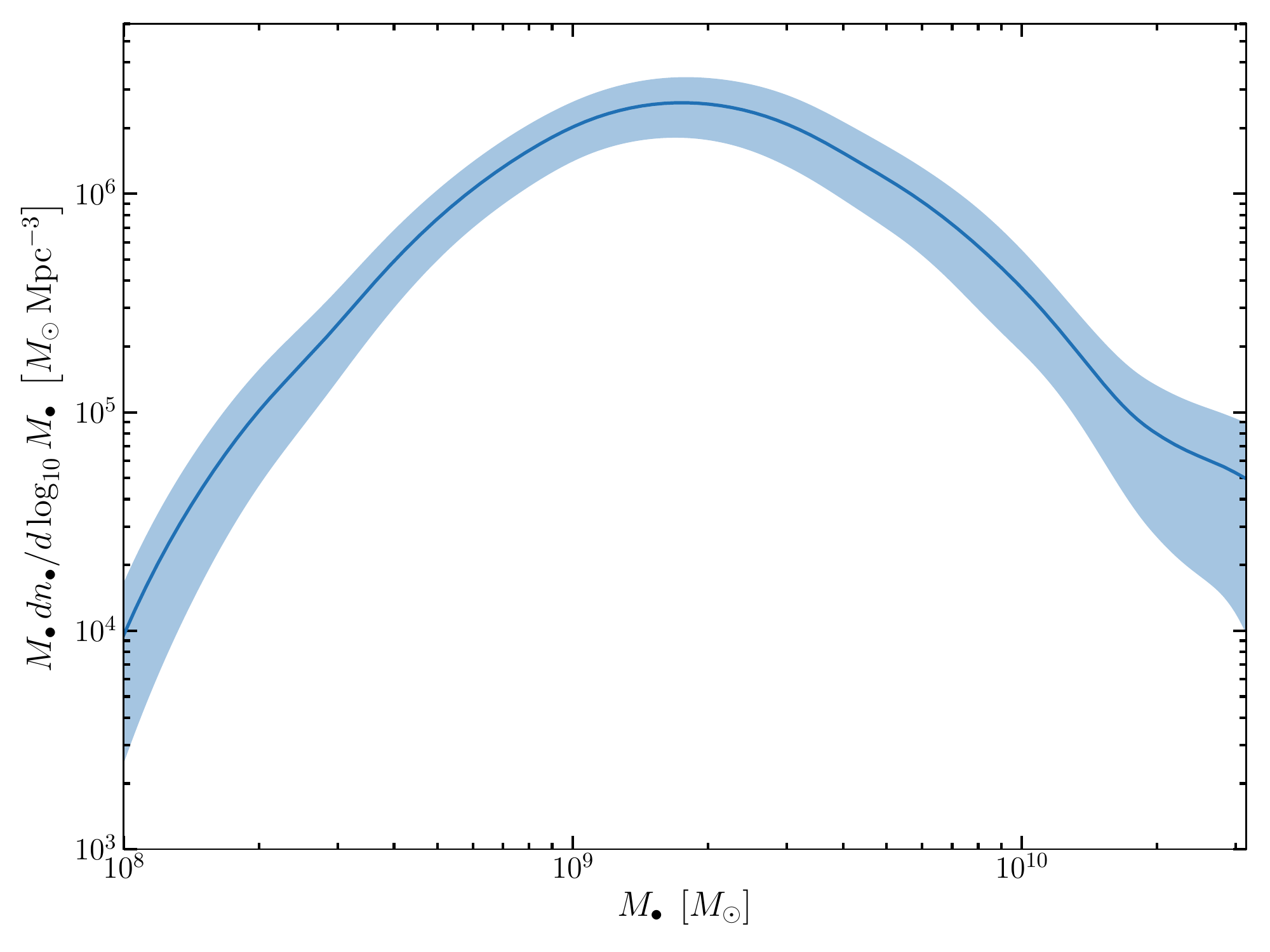}
    \caption{The local BH mass function as predicted from the SDSS data, assuming a fiducial value for the duty cycle $f_{\rm opt}=2\times10^{-3}$. The shaded band represents a 1$\sigma$ confidence interval. In the mass range $\Mblack\lesssim 10^9\Msun$, the above BHMF under predicts the actual BHMF due to flux limitations of the SDSS.  \label{fig:BHMF}}
\end{center}
\end{figure}

The QMF, $\Phi_Q$, estimated directly from the SDSS data, is defined as the number of quasars (active and luminous galactic nuclei) per co-moving volume per log BH mass interval
\footnote{Throughout this Letter we use log base 10.}
\begin{equation}
    \Phi_Q(\Mblack,z)=\frac{d^2N_Q}{dV_C(z)\;d\log \Mblack},
\end{equation}
where $V_C(z)$ is the co-moving volume of a sphere of co-moving radius $D_C(z)$.  As mentioned in Sec. \ref{sec:data}, we are dealing with a spherical region within which we find a constant co-moving quasars number density.  Estimating the QMF is therefore achieved by calculating the mean number of quasars seen by a random observer in a given mass interval and dividing it by the observed volume. Our selection of random observers cover a spherical region of radius $1000\,\rm Mpc$, each observes a maximal distance of $1300\,\rm Mpc$ (corresponding to a redshift $z\simeq0.32$). See the Appendix for additional details on the method used for this calculation.


The local BHMF, $\Phi$,  is defined as the number of BHs, active or dormant, per co-moving volume per log BH mass interval. The BHMF is related to the QMF via
\begin{equation}
  \Phi_Q=f_{\rm opt}\Phi,
\end{equation}
where $f_{\rm opt}$ is the fraction of black holes that shine as optically luminous quasars at a given time. We adopt $f_{\rm opt}$ as calculated in ~\citet{2010MNRAS.406.1959S}. There, based on~\citepos{2009ApJ...697.1656S} measurements of the luminosity-dependent clustering of quasars in SDSS DR5~\citep{2000AJ....120.1579Y,2007ApJS..172..634A,2007AJ....134..102S}, $f_{\rm opt}$  is derived by assuming a mean relation between quasar luminosity and its host halo mass, and matching the co-moving number density of quasars to that of halos, corrected for the duty cycle. For our interest we shall use a fiducial, somewhat conservative, value of $f_{\rm opt}\simeq 2\times 10^{-3}$, which~\citet{2010MNRAS.406.1959S} find to be consistent with both their analysis and continuity equation modelling of the BH population~\citep{1971ApJ...170..223C,1992MNRAS.259..725S,2009ApJ...690...20S}. The resulting local BHMF is shown in Fig.~\ref{fig:BHMF}. The drop of the mass function at masses above $\Mblack\sim10^9\Msun$ is physical \citep{Kelly_2012}, the suppression below $\Mblack\sim10^9\Msun$ is attributed to the flux limitation selection explained earlier in this section.

\subsection{The Distribution of Black Holes of a Given Angular Size}
We define the  angular size of a BH shadow, $\theta_{\rm shad}$,~\citep{Johannsen:2010ru,Psaltis:2018xkc}, and gravitational influence angular size, $\theta_{\rm grav}$, \citep{Peebles_1972} as\footnote{To be precise, the angular size of the BH shadow is given by $A G \Mblack/c^2D_A$, where $A$ is a BH spin dependent number, consistent with 5 to within 4\%~\citep{Johannsen:2010ru,Psaltis:2018xkc}.}
\begin{equation}\label{eq:theta_BH}
    \theta_{\rm shad}=\frac{5GM_\bullet}{c^2D_A(z)}\;\;,\;\;\theta_{\rm grav}=\frac{G \Mblack}{\sigma_v^2D_A(z)} \, ,
\end{equation}
respectively, where $D_A(z)$ is the angular diameter distance to a BH at redshift $z$ and $\sigma_v$ is the velocity dispersion around the BH, for which we phenomenologically assume the $\Mblack- \sigma$ relation from~\citet{Kormendy:2013dxa}. Whenever possible we will address the two angles collectively as $\theta_\bullet$. Starting from a BHMF, the differential number of BH seen with angular size $\theta_\bullet>\theta$ is given by
\begin{equation}
    d N_{\bullet}(\theta)=\Theta\left(\theta_{\bullet}-\theta\right) \Phi(M_{\bullet}) d V_{C}(z) d \log M_{\bullet},
\end{equation}
where $\Theta$ is the Heaviside function. Since each random observer in our analisys observes a small redshift interval, we have explicitly dropped the BHMF redshift dependence in the equation above. Integrating this differential over $V_C$, $M_\bullet$ or both, results in the number of BHs with angular size larger than $\theta$, per log mass, per co-moving volume, or in total, respectively. Since the luminosity of distant objects drops as $(1+z)^{-4}$, we will integrate the mass function only up to $z_{\rm max}$ so that we do not count sources that are too faint. As mentioned earlier, in our analysis $z_{\rm max}\simeq 0.32$, within this redshift range $D_A(z)$ is a monotonically increasing function of $z$, therefore $z$ is determined uniquely for a given $D_A$. Under these restrictions we find

\begin{align}
    &\frac{d N_{\bullet}\left(\theta_{\bullet}>\theta\right)}{d \log M_{\bullet}}=V_C(\bar{z})\Phi(M_{\bullet})\;\label{eq:dN_BH_theta}\\
    &\frac{d N_{\bullet}\left(\theta_{\bullet}>\theta\right)}{d  V_{C}}= \int_{M_{\min }}^{\infty} \Phi(\Mblack) d\log\Mblack\label{eq:dN_BH_theta_V}
\end{align}
where $\bar{z}=\min[z_{\rm max},z_\theta]$ such that $D_A(z_\theta)$ satisfies $\theta_\bullet=\theta$ for a fixed $\Mblack$, while $M_{\rm min}$ satisfies $\theta_\bullet=\theta$ for a fixed $D_C$ (see. Eq.~\ref{eq:theta_BH} and bear in mind that $\sigma_v^2$ is a function of the BH mass.) The total number of BHs can be obtained by integrating Eq.~\eqref{eq:dN_BH_theta} over $\Mblack$, or Eq.~\eqref{eq:dN_BH_theta_V} over $V_C$.

\begin{figure*}
\begin{center}
    \includegraphics[width=0.48\linewidth]{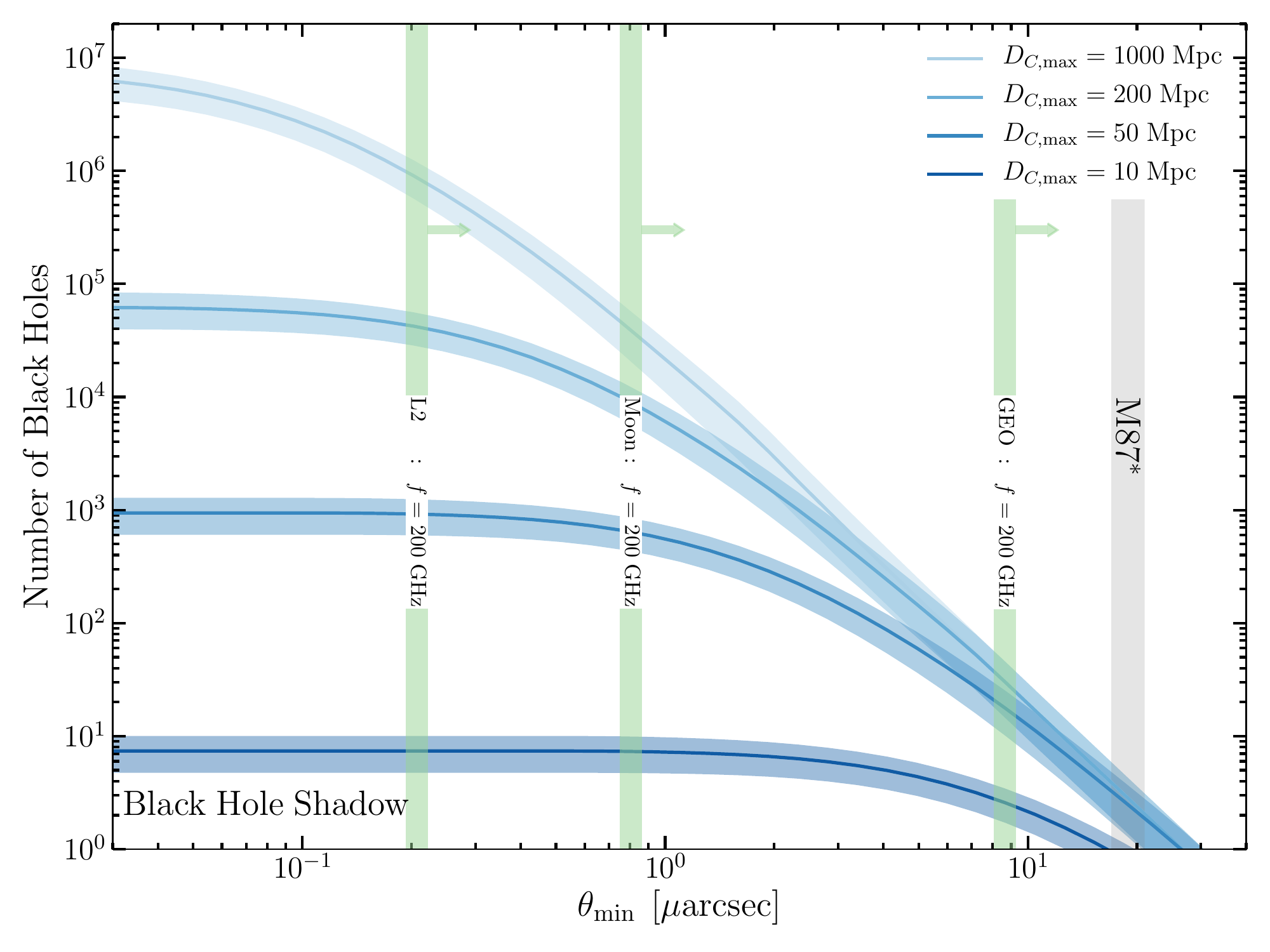}\;\;\;\;\includegraphics[width=0.48\linewidth]{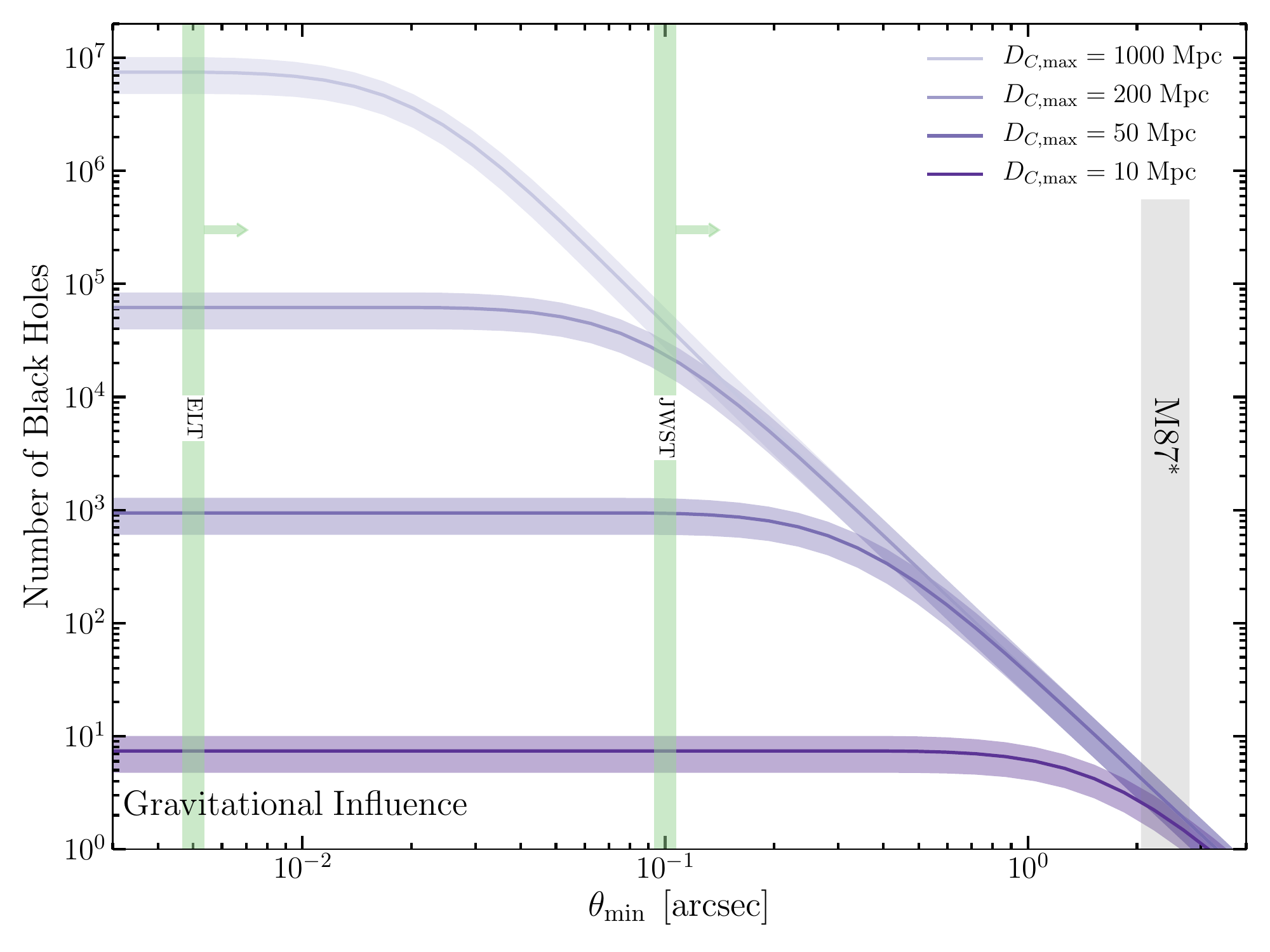}
    \caption{The cumulative number of BHs a telescope with angular resolution $\theta_{\rm min}$ is predicted to to resolve, for different maximal accessible distance ranging in 10 - 1000 Mpc. The \textbf{shaded blue/purple bands} represent the $1\sigma$ confidence interval. \textit{Left:} Prediction for BH shadow searches. The \textbf{green vertical bands} indicate, from right to left, the angular resolution of a hypothetical future VLB observation (at frequency of $200$ GHz) with one  Earth-based receiver and the other one in geostationary orbit, on the Moon and at the Sun-Earth Lagrange point L2, respectively. \textit{Right:} Prediction for gravitational influence searches, here the \textbf{green vertical bands} indicate, from right to left, the angular resolution of the JWST and the future ELT \citep{JWST, ELT_2005}. On both panels \textbf{grey bands} shows for reference the angular size of $\mathrm{M87^*}$, using its shadow size as reported in~\citetalias{EHT} and the velocity dispersion of $\mathrm{M87}$ as reported in~\citet{2014ApJ...785..143M}. }
   
\label{fig:Total_No}
\end{center}
\end{figure*}

\section{Results}
In Fig.~\ref{fig:Total_No} we show the main result of this study: the expected number of BHs that can be resolved by a telescope with a given angular resolution. In both panels we demonstrate the dependence of our result on the maximal accessible distance, which is limited by the telescope sensitivity. For reference, we show on both panels the angular size of $\mathrm{M87}^*$, for which \citetalias{EHT} reported a shadow of angular dimension $\theta_{\rm shad}=19\pm2\;\mu\rm arcsec$. Since $\theta_{\rm grav}={c^2}\theta_{\rm shad}/5 \sigma_v^2$, we may use the EHT result together with the velocity dispersion of $\mathrm{M87}$ reported in~\citet{2014ApJ...785..143M} to calculate the angular size of the gravitational influence region of $\mathrm{M87}^*$. We find this to be $\theta_{\rm grav}=2.40^{+0.39}_{-0.35}\;\rm asrcsec  $. As seen in Fig.~\ref{fig:Total_No}, our predictions are consistent with $\mathcal{O}\sim1$  BHs with dimensions similar to those of $\mathrm{M87}^*$ accessible to current day experiments. To appreciate the number oh BHs next-generation missions could observe, we show on the left panel of Fig.~\ref{fig:Total_No} the angular resolution of a VLB observatory at frequency of $200$ GHz, with one receiver located on earth and the other in a geostationary orbit, on the Moon or in the Sun-Earth Lagrange point L2. Such VLB observatories are further motivated as they offer novel and exciting tests of general relativity~\citep{2020SciA....6.1310J,2020PhRvD.102l4004G}. On the right panel we show the resolution of the upcoming JWST and the ELT~\citep{JWST, ELT_2005}. 

To gain some qualitative intuition we analyze here the small and large $\theta_{\rm min}$ behaviour of Fig.~\ref{fig:Total_No}. The plateau at small resolution is the result of the finite number of BHs contained within a sphere of radius $D_{C,\rm max}$, we find it to be
\begin{equation}\label{eq:assy_theta}
     N_\bullet(\theta_\bullet>0)\simeq10^3 \left(\frac{2\times10^{-3}}{f_{\rm opt}}\right)\left(\frac{D_{C}}{50\;\rm Mpc}\right)^3,
\end{equation} 
where we have re-introduced the dependence on $f_{\rm opt}$ for completeness. This number would in practice be larger when the selection of BHs lighter than $10^9\Msun$ is corrected for.

The tail at large resolution on the other hand demonstrates the scaling $N_\bullet\sim \theta_{\rm min}^{-3}$. There, as a result of the fact that at large resolution we observe only nearby BHs, $D_C\simeq D_A$ and one has
\begin{align}
    N_\bullet(\theta_\bullet>\theta_{\min})&\simeq \frac{4 \pi G^3}{3\theta_{\rm min}^{3}}\int_{M_{\min}}^\infty S^3(\Mblack)\Mblack^3\Phi(\Mblack) d\log \Mblack\nonumber\\ 
    &\simeq 
    \begin{cases}
    \left(\frac{27\;\rm \mu arcsec}{\theta_{\rm min}}\right)^3&\rm shad\\
    \left(\frac{3.3\;\rm arcsec}{\theta_{\rm min}}\right)^3&\rm grav
    \end{cases}\;,
\end{align}
where $S=5c^{-2}$ for the BH shadow and $S=\sigma_v^{-2}$ for the gravitational influence.

\begin{figure*}
\begin{center}
    \includegraphics[width=0.48\linewidth]{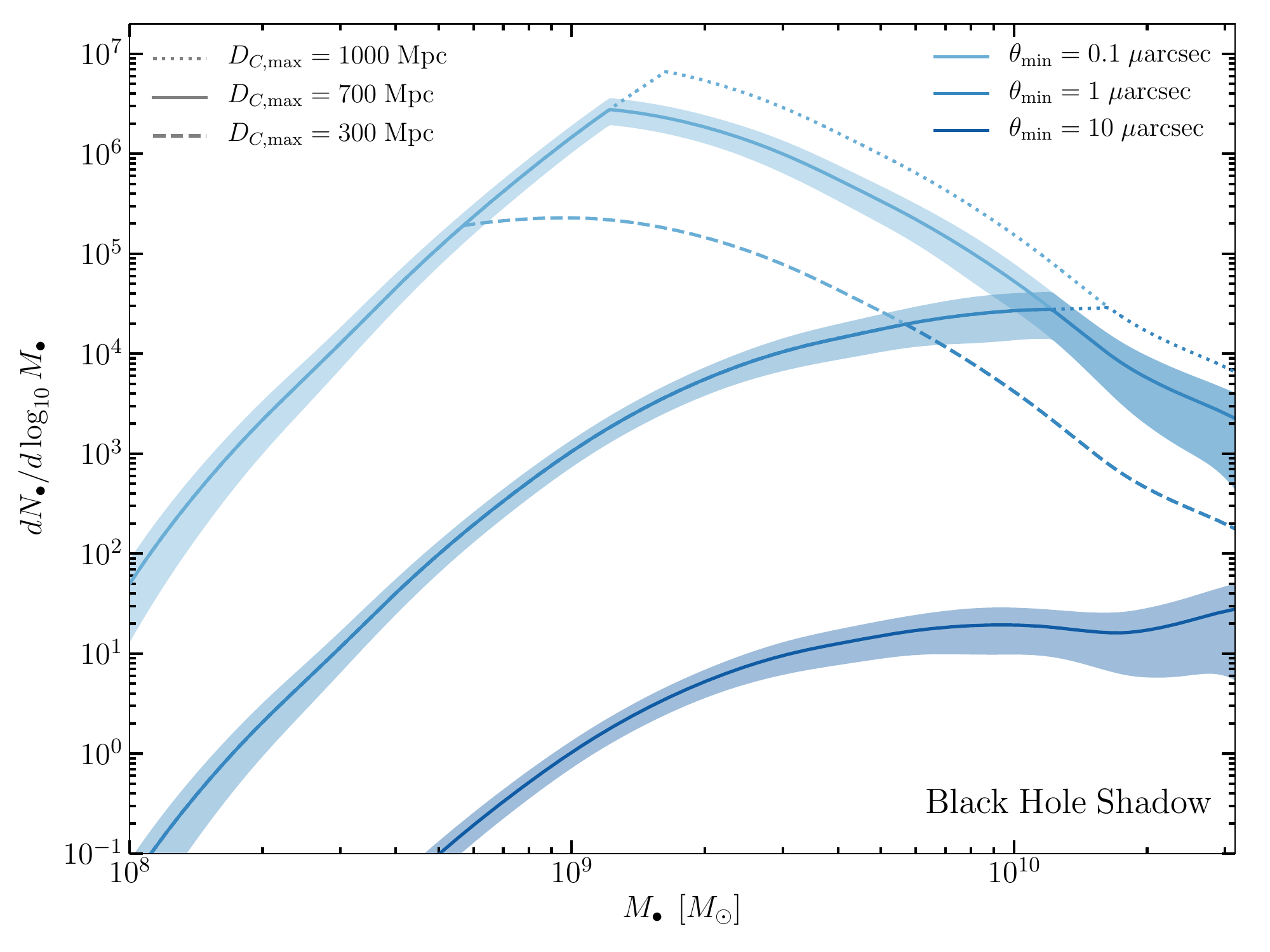}\;\;\;\;\includegraphics[width=0.48\linewidth]{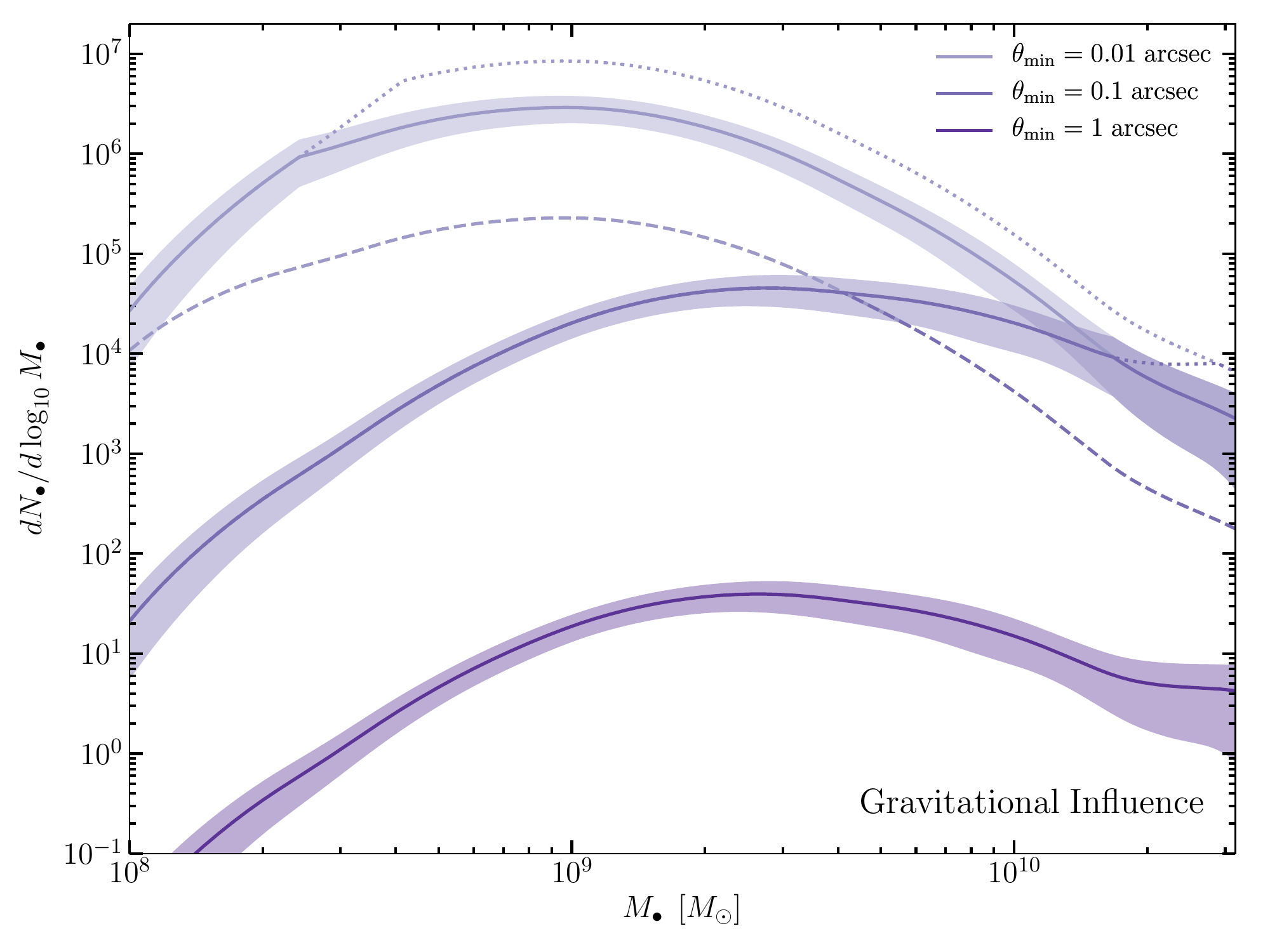}
    \caption{Predicted number of BHs of angular dimension larger than a given $\theta_{\rm min}$, per log mass interval. \textbf{Solid lines} shows the result assuming a maximal accessible co-moving distance of 700 Mpc. The \textbf{shaded blue/purple bands} represent the $1\sigma$ confidence interval around the solid lines. \textbf{Dashed and Dotted lines} demonstrate the sensitivity of our result to the maximal distance that a given telescope can observe. We omit the confidence interval from the dashed and dotted lines for clarity. The sharp feature appears because of two different causes that limit the amount of observables BHs; To the left of the sharp feature the number of BHs is limited by resolution, while to the right it is limited by the maximal observable distance. See also main text.}
   
\label{fig:Mass_Function}
\end{center}
\end{figure*}

\begin{figure*}[t]
    \begin{center}
    \includegraphics[width=0.48\linewidth]{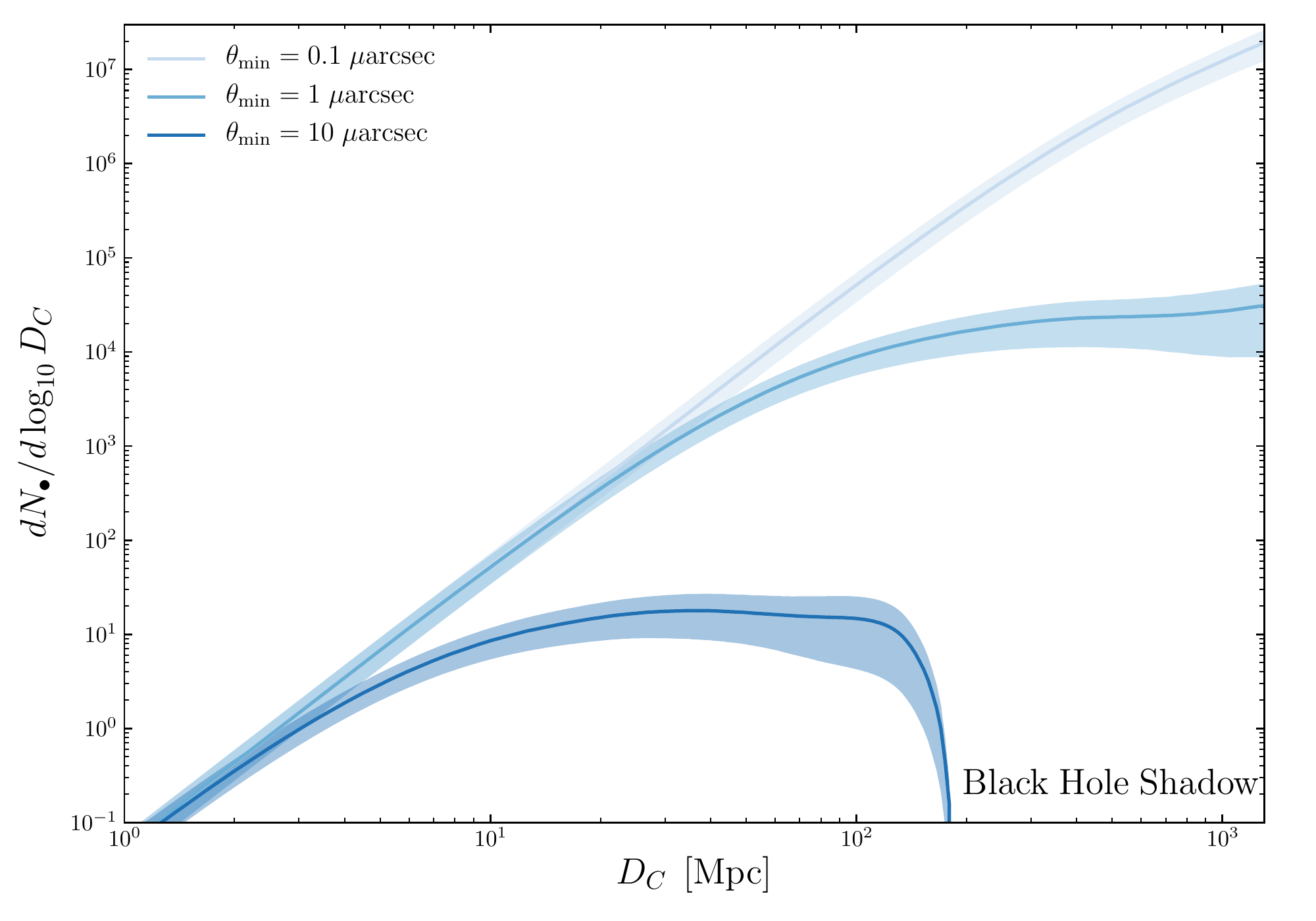}\;\;\;\;\includegraphics[width=0.48\linewidth]{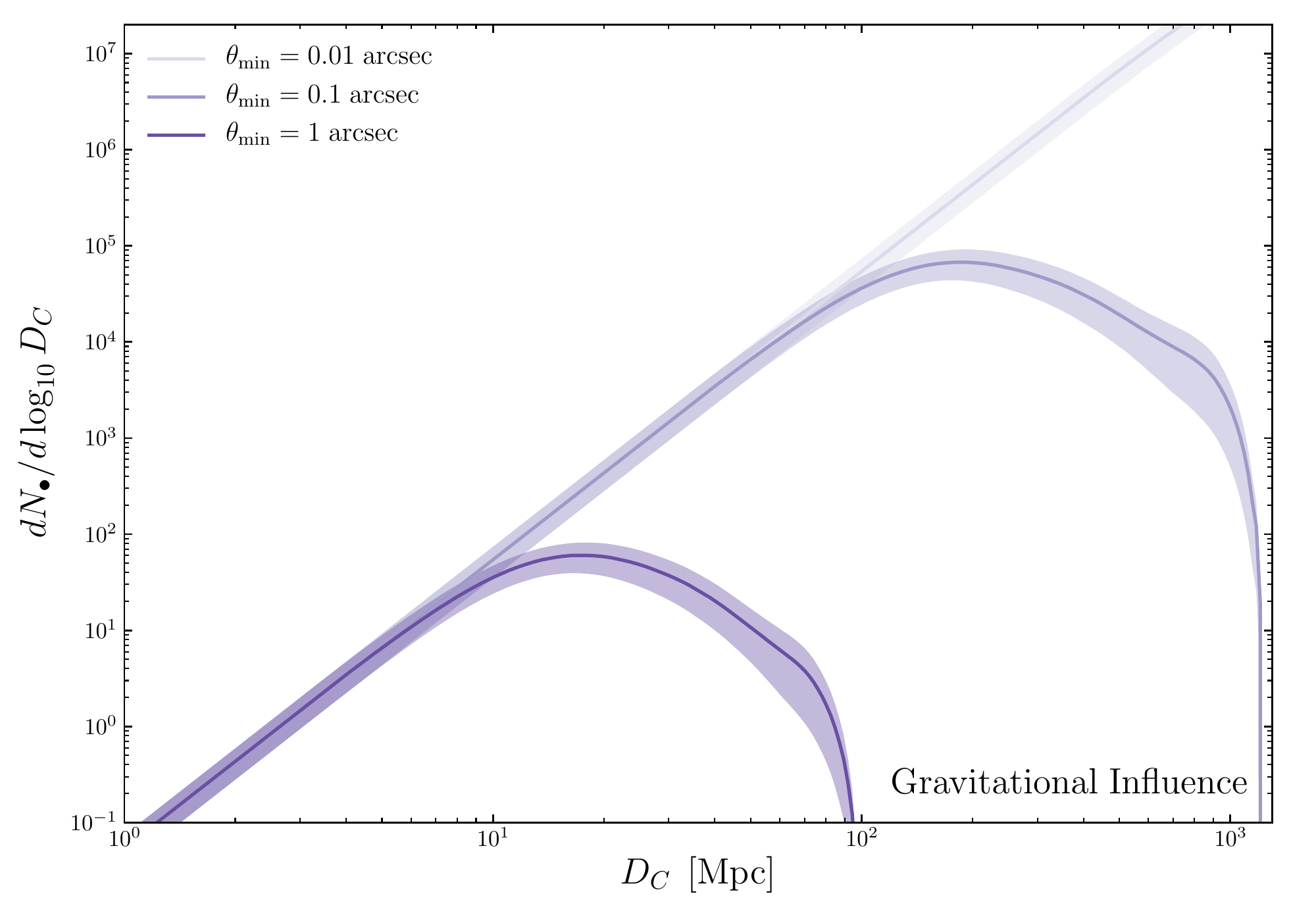}
    \caption{Predicted number of BHs of angular dimension larger than a given $\theta_{\rm min}$, per log comoving distance interval. The \textbf{shaded blue/purple bands} represent the $1\sigma$ confidence interval around the solid lines.}
   
    \label{fig:Dist_Function}
    \end{center}
\end{figure*}

In Fig.~\ref{fig:Mass_Function} we show the total number of BH expected to be seen with a telescope of given resolution, per unit log mass and as a function of the BH mass. To demonstrate the effect of the telescope sensitivity we show (in dotted and dashed lines) how the curve changes when the maximal accessible distance $D_{C,\max}$ is changed. At the high-mass-end of the curve the dependence is simply $N_\bullet\sim D_{C,\rm max}^3$. The sharp features in the curves are a result of the definition of $\bar{z}=\min[z_{\rm max},z_\theta]$ (see Eq.~\ref{eq:dN_BH_theta}); physically, to the left of the kink the relevant volume is limited by resolution $\theta_{\rm min}$, and to the right by the maximal distance $D_{C,\rm max}$. 
We stress again that for $\Mblack\lesssim10^9\Msun$  the reported results underestimate the actual density of BHs due to the SDSS flux limitation.

Last, in Fig.~\ref{fig:Dist_Function} we show the total number of BH expected to be detectable by a telescope of  a given resolution, per unit log distance and as a function of co-moving distance. The small distance behaviour of those plots is simply given by the derivative of Eq.~\eqref{eq:assy_theta} with respect to $D_C$. The drop at large distances is due to the fact that the further away you observe the more massive the BH needs be in order to be observable for a given resolution, together with the drop of the BHMF at mass heavier than$~10^9\Msun$.

\section{Discussion and Conclusion}
Using a statistical approach based on data from the SDSS, we have inferred the local abundance of BHs heavier than $10^8 \Msun$. We have shown that future VLB interferometry observations with longer arm lengths will be able to image scales comparable to the event horizon of many more BHs. In addition, IFS observations with the upcoming JWST and ELT will be able to dynamically measure the mass locked into BHs within many local galactic nuclei \citep{JWST}.

We conclude by stressing some of the underlying assumptions and caveats of our analysis.
First, a word of caution regarding BH mass estimates is warranted. BH mass estimates in the SDSS DR7 are performed via virial mass estimators, based on the emission lines $\rm H\beta$, $\rm MgII$, and $\rm CIV$. These mass estimates are affected by uncertainties, which are possibly large, especially at the high-luminosity end. As shown by \cite{Shen_2008}, the observed virial BH mass distribution for the SDSS sample used in this study is biased by a factor of $\sim 3$, towards the high end of the BH mass distribution. A factor $\sim 2-4$ uncertainty is typical for mass estimates of high-$z$ quasars (see, e.g., \citealt{Unal_2020}). While this bias does not significantly affect the results in the present study, the reader should be cautioned that the very high-mass BHs reported in this statistical study could possibly be spurious. For completeness, in Appendix~\ref{sec:effect_of_sdss_bias} we study more quantitatively the effect of this bias on the robustness of our results. Fig.~\ref{fig:Total_No_app} in that Appendix shows a ``worst case scenario'' version of Fig.~\ref{fig:Total_No}, where all BH mass in the data used are scaled-down by a factor of 3.

Our analysis is based on the duty cycle as estimated in~\citet{2010MNRAS.406.1959S}. A more dedicated study of the duty cycle and its dependence on the BH mass (see, e.g., \citealt{DeGraf_2017}) can improve the accuracy of our analysis. As we adopted a single value for $f_{\rm opt}$, a strong mass dependence could significantly affect our results. 

Another assumption we have taken was that the BHMF calculated at $z\sim1.4$ is a good estimate of the BHMF at $z\sim0$. Accounting for the evolution of the BHMF from $z=1.4$ to $z=0$ should only increase the number of observed BHs, since the BHs could only grow (in mass and therefore in size) during this time. However, due to the activity drop at $z\lesssim 1$ we expect this effect to be rather small.

As a final note, we wish to emphasize again that due to SDSS flux limitation, our analysis under-predicts the number of BHs lighter than $\sim 10^9\Msun$. By means of complementing data from the SDSS with other surveys, the resulting number of BHs that are expect to be observed in the local Universe could be higher. 

In spite of the fact that a large fraction of central BHs in the local Universe may be dormant and dark, the future is bright. In this study we showed that forthcoming observations with VLB techniques, as well as IFS analysis with the JWST and the ELT will likely greatly expand our census of the local population of super-massive BHs.

\section*{Acknowledgements}
We thank Melanie Habouzit, Simon Knapen, Siddharth Mishra-Sharma, Yonatan Piasetzky, Hagai Rossman, Omer Shabtai and Yue Shen for useful discussions. N.J.O. is grateful to the Azrieli Foundation for the award of an Azrieli Fellowship, and is partially supported by the European Research Council (ERC) under the EU Horizon 2020 Programme (ERC-CoG-2015 - Proposal n. 682676 LDMThExp). F. P. acknowledges support from a Clay Fellowship administered by the Smithsonian Astrophysical Observatory and from the Black Hole Initiative at Harvard University, which is funded by grants from the John Templeton Foundation and the Gordon and Betty Moore Foundation.

\bibliography{main}

\appendix

\section{The Effect of the SDSS High Luminosity Bias} 
\label{sec:effect_of_sdss_bias}
\begin{figure*}
\begin{center}
    \includegraphics[width=0.48\linewidth]{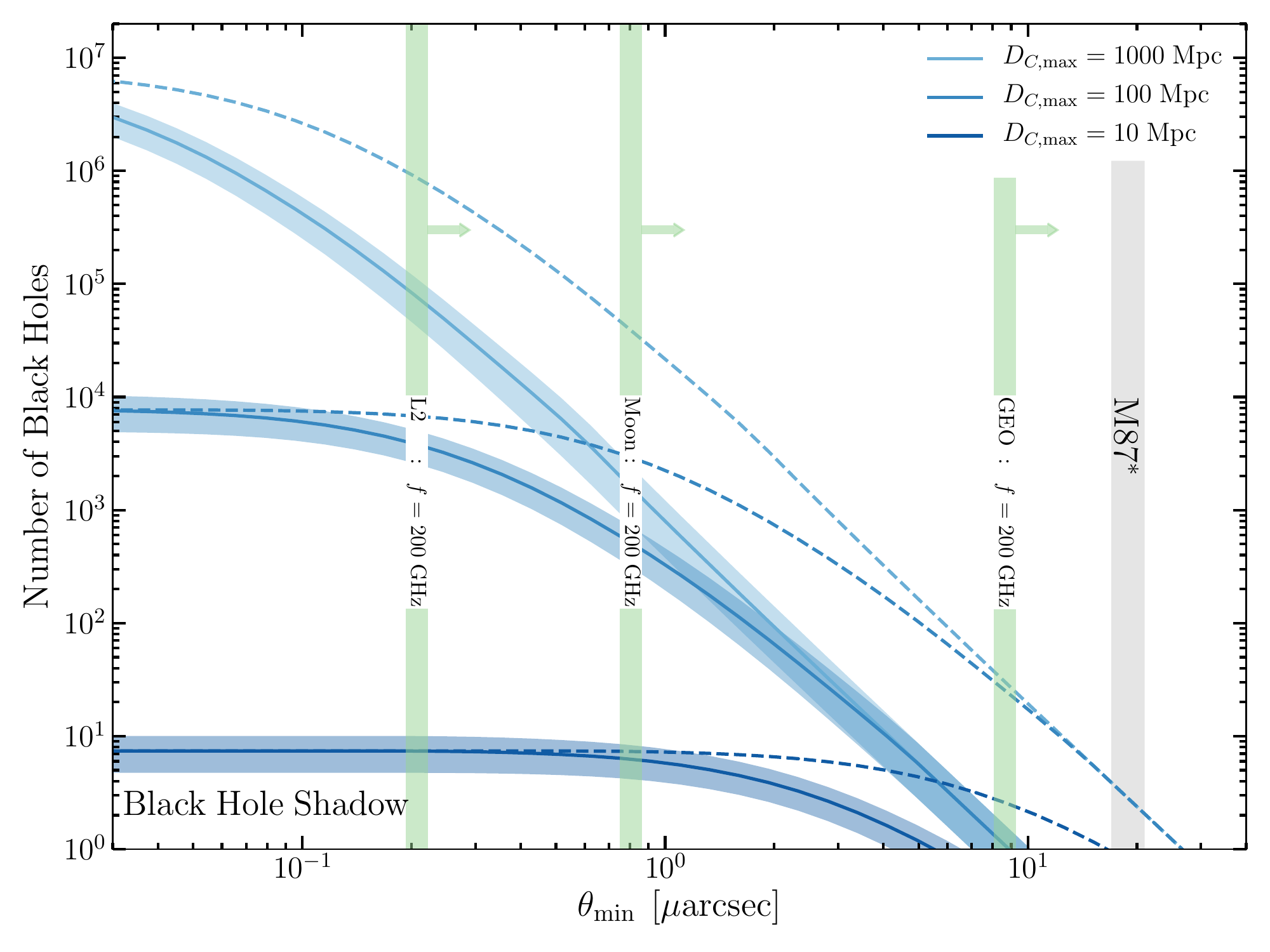}\;\;\;\;\includegraphics[width=0.48\linewidth]{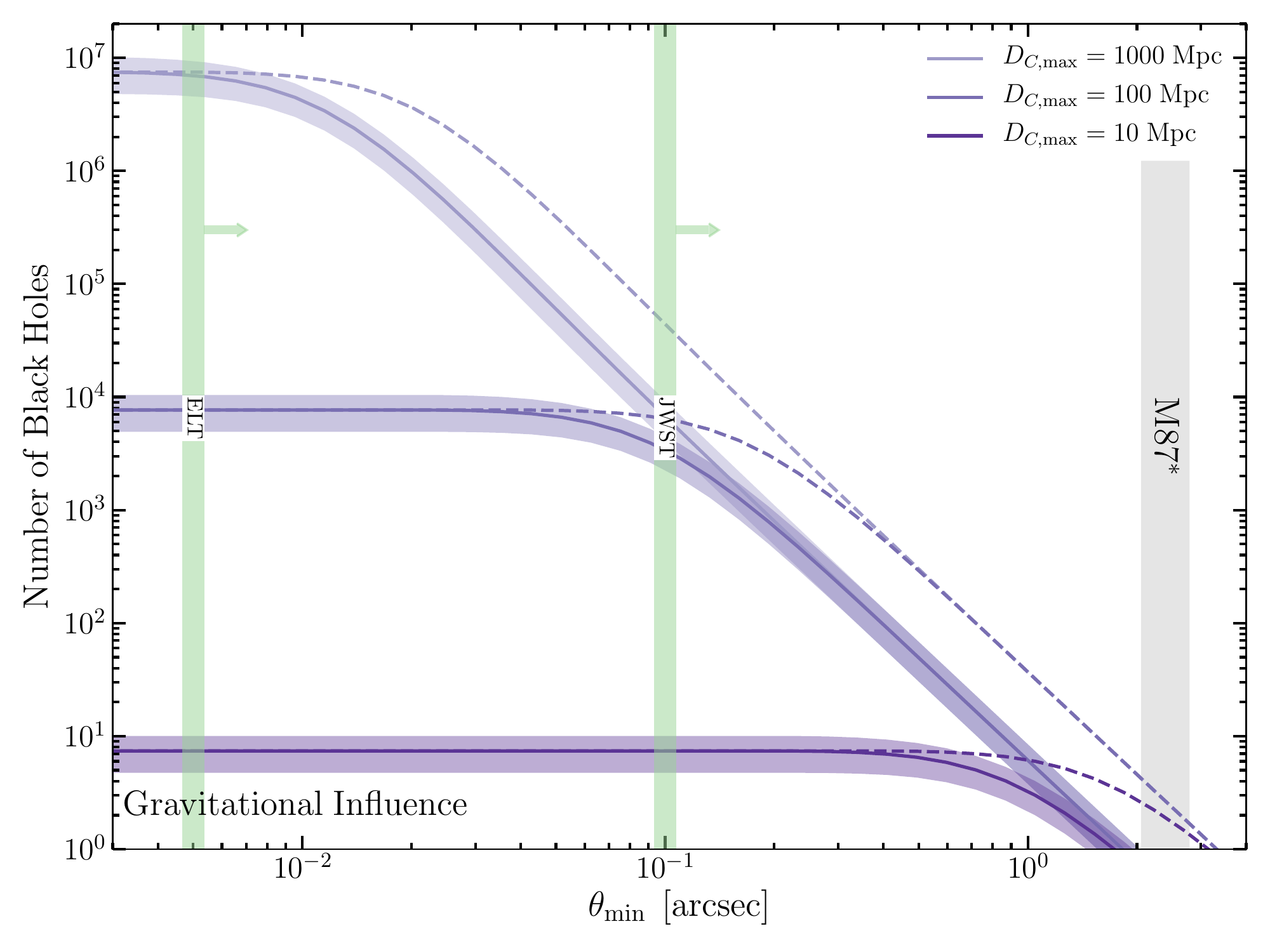}
    \caption{The number of BHs a telescope with angular resolution $\theta_{\rm min}$ is predicted to to resolve, for different maximal accessible distance ranging in 10 - 1000 Mpc. \textbf{Solid lines and $\boldsymbol{1\sigma}$ bands} were calculated assuming a constant scaling by a factor 3 of all BH mass used in the analysis. \textbf{Dashed lines} shows the unscaled results.  See the caption of Fig.~\ref{fig:Total_No} for further details.}
   
\label{fig:Total_No_app}
\end{center}
\end{figure*}

In this short appendix we wish the estimate the sensitivity of our results to the BH mass bias as studied in \citet{Shen_2008}, where  a maximal bias of $~3$ reported. This bias is not easily corrected for, however, to estimate the sensitivity of our results to that bias, here we shall treat it as universal bias and scale-down all the BH masses used in our analysis by a factor of 3. The effect on the reported BHMF is a trivial shift of the BHMF by a factor of 3 to the left (see. Fig.~\ref{fig:BHMF}).  The effect on the total number of BH seen with angular size larger than $\theta_{\rm min}$ is slightly more involved, the results are shown in Fig.~\ref{fig:Total_No_app}, which should be compared with the un-scaled version shown in Fig.~\ref{fig:Total_No}. The dashed lines in fig.~\ref{fig:Total_No_app} shows for clarity the un-scaled results. Qualitatively, the mass scaling results in a scaling of the $\theta_{\rm min}$ axis by a factor of $3$ in the BH shadow calculation, and by a factor of $\sim 1.82$ in the gravitational influence calculation. The different scaling is a result of the additional BH mass dependence arising from the $\Mblack- \sigma$ relation in the gravitational influence calculation \citep{Shankar_2019}. We therefore conclude that at a fix $\theta_{\rm min}$ the total number of BHs that can be resolved is diminished at most by a factor of $29$ for BH shadow searches, and at most by a factor of $\sim 6$ for the gravitational influence searches.

\section{Geometrical Determination of the Observed Number of Quasars} 
Usually a Monte Carlo (MC) process is sufficient to estimate the distribution of objects in 3D space. One chooses a large number of randomly positioned observers and calculate the distribution of those objects as seen by each observer. The "distribution of distributions" is therefore sampled by the randomly located observers. Here we aim to study the distribution of quasars (and infer that of BHs) with a sufficiently large angular size. To achieve this, we would have to sample our observed volume to a very fine resolution, making the MC process very computationally expensive. To exemplify, in order to probe quasars with of mass~$\sim 10^9\;M_\odot$ and shadow with angular size~$\sim5\;\mu\rm{arcsec}$, we would have to use more than $10^7$ observers. 

To bypass this difficulty we devise a geometric method for calculating the mean and standard deviation of the number of quasars seen by a random observer with an angular size larger than $\theta$. Each quasars in the SDSS catalogue is assigned with a co-moving position vector $\vecb{r}_i$ and a mass $M_i$. The number of quasars seen by an observer located at $\vecb{r}$, with  distance smaller than $r_{\rm max}$ is given by
\begin{equation}
    N(\vecb{r})=\sum_i\int_{r'<D_{C,\rm max}} d^3r'\delta^{(3)}(\vecb{r}+\vecb{r}'-\vecb{r}_i).
\end{equation}
The mean number of quasars is then obtained by averaging the above expression over the observed volume $V_O$
\begin{equation}
    \bar{N} =\frac{1}{V_O}\int_{V_O}d^3r N(\vecb{r}).
\end{equation}
Since all points closer than $D_{C,\rm max}$ to $\vecb{r}_i$ form a ball of radius $D_{C,\rm max}$ centred at $\vecb{r}_i$, which we denote $B_i$, the above integral is simply given by
\begin{equation}
   \bar{N} =\sum_i \frac{V_{O\cap B_i}}{V_O}, 
\end{equation}
i.e., the sum of the volume of the intersection of a ball $B_i$ and the entire observed volume. With similar steps, we find that the variance is given by
\begin{equation}
    {\rm Var}[N]=2\sum_{j>i}\frac{V_{O\cap B_i\cap B_j}}{V_O}+\bar{N}_\theta-\bar{N}_\theta^2,
\end{equation}
where now the intersection is of two balls and the observed region. Higher cumulants can be calculated analogously and require the calculation of volume of intersection of a larger number of balls.  We have checked this calculation against MC when the MC converges and the two methods agree.     

As described in Sec. \ref{sec:data}, we restrict our analysis to spherical observed volumes. One reason for this choice is that the volume of intersection of two and three spheres has been studied in the literature and is given in closed form, allowing for an efficient calculation~\citep{doi:10.1021/j100299a035,Chkhartishvili2001,Petitjean2013}. To avoid biases at the edges of our observed volume, we shall always use observed sphere of radius $R_O$ satisfying $R_O+r_{\rm max}\leq 2300 \,\rm Mpc$, such that all quasars influencing our observed region is within our data set.

\end{document}